\documentclass{amsart}
\usepackage{amsmath, amsthm, latexsym, amssymb, amsfonts}\usepackage[pdftex]{graphicx,color} 

\newenvironment{pf}{\proof[\proofname]}{\endproof}
\theoremstyle{plain}
\newtheorem{Th}{Theorem}[section]

\newtheorem{Cor}[Th]{Corollary}
\newtheorem{Prop}[Th]{Proposition}

\numberwithin{equation}{section}
\numberwithin{figure}{section}
\theoremstyle{definition}
\newtheorem{Rem}[Th]{Remark}

\newtheorem{Def}[Th]{Definition}

\newcommand{\cal}[1]{\mathcal{#1}}

\newcommand{\Z}{\mathbb Z}
\newcommand{\R}{\mathbb R}
\newcommand{\F}{\mathbb F}

\newcommand{\D}{\Delta}
\newcommand{\cA}{\cal A}

\newcommand{\cC}{\cal C}

\newcommand{\cL}{\cal L}

\newcommand{\cP}{\cal P}

\newcommand{\T}{{\mathbb T}}

\newcommand{\K}{{\mathbb K}}

\newcommand{\spn}{\operatorname{span}}

\newcommand{\GL}{\operatorname{GL}}
\newcommand{\AGL}{\operatorname{AGL}}

\newcommand{\rs}[1]{Section~\ref{S:#1}}

\newcommand{\rp}[1]{Proposition~\ref{P:#1}}

\newcommand{\re}[1]{(\ref{e:#1})}
\newcommand{\rc}[1]{Corollary~\ref{C:#1}}
\newcommand{\rt}[1] {Theorem~\ref{T:#1}}
\newcommand{\rd}[1]{Definition~\ref{D:#1}}
\newcommand{\rf}[1]{Figure~\ref{F:#1}}

\begin{document}


\title{Lattice polytopes in coding theory}
\author[Ivan Soprunov]{Ivan Soprunov}
\address[Ivan Soprunov]{Department of Mathematics\\ Cleveland State University\\ Cleveland, OH USA}
\email{i.soprunov@csuohio.edu}
\thanks{The author is partially supported by NSA Grant H98230-13-1-0279}


\begin{abstract} 
In this paper we discuss combinatorial questions about lattice polytopes
motivated by recent results on minimum distance estimation for toric codes. We also
include a new inductive bound for the minimum distance of generalized toric
codes. As an application, we give new formulas for the minimum distance of 
generalized toric codes for special lattice point configurations.
\end{abstract}

\maketitle


\section*{Introduction}

Toric codes are examples of a large class of evaluation codes 
studied by Goppa, Tsfasman, Vl\v{a}dut, and others, using methods of algebraic geometry \cite{TVN}.
Yet the construction is very explicit: Given a lattice polytope $P$ in $\R^m$,  consider the set of all 
$m$-variate polynomials whose exponent vectors lie in $P$. 
The code is produced by evaluating these polynomials at the points of $(\F_q^*)^m$. 
This makes toric codes a wonderful example of an interconnection between algebraic geometry (toric varieties), geometric combinatorics (lattice polytopes), and coding theory.
Toric codes were first introduced by J.~Hansen in \cite{Ha1} for $m=2$ 
and have been actively studied in the last decade. Here is a list of some recent
papers on the subject: \cite{Ha2, Jo, LSc, LSch, Ru, SoSo, SoSo2, UmVe}.
Apart from numerous theoretical results,  about a dozen new ``champion" toric codes and generalized toric codes have been found just recently \cite{Lit11, BrKasp13, BrKasp-seven}.  A ``champion" code is the one that has the largest known minimum distance for a given  block length and dimension, as in the table of best known codes \cite{Gra}.

In this paper we concentrate on combinatorial questions about lattice polytopes which arise when one
studies the minimum distance of toric codes. In \rs{TC} we relate the minimum distance to a geometric invariant
called the  Minkowski length of $P$. In particular, we look at the problem of estimating the number of
lattice points in polytopes of fixed  Minkowski length. \rs{GTC} is concerned with generalized toric codes. 
There we prove a general inductive bound for the minimum distance. As an application we generalize
previously known formulas for the minimum distance (\rt{SoSo2}) to generalized toric codes. 
In addition, we present two recently found champion codes.    

\section{Preliminaries}\label{S:Prel}

\subsection{Linear Codes} To set our notation we start with basic definitions from coding theory. 
Throughout the paper, $\F_q$
denotes a finite field of $q$ elements and $\F_q^*$ its multiplicative group of non-zero elements.
A subspace $\cC$ of $ \mathbb{F}_{q}^{n}$ is called a \textit{linear code}, and its elements 
${c}=(c_{1},\dotsc,c_{n})$ are called \textit{codewords}.  The number $n$ is called the {\it block length} of $\cC$.
The {\it weight} of $c$ in $\cC$ is the number of non-zero entries in $c$.
The {\it distance} between two codewords $a$ and $b$ in $\cC$ is the weight of $a-b\in\cC$.
The minimum distance between distinct codewords in $\cC$ is the same as the minimum weight of  non-zero codewords in $\cC$.
The block length $n$, the dimension $k=\dim(\cC)$, and the minimum
distance $d=d(\cC)$ are the parameters of $\cC$. A code with parameters $n$, $k$, and $d$ is referred to as an $[n,k,d]_{q}$-code.

\subsection{Newton polytopes} 
Let $f$ be a polynomial in $m$ variables over a field $\K$. If we allow negative exponents in
the monomials of $f$ we call it a Laurent polynomial. The set of the exponent vectors of the
monomials appearing in $f$ is called the {\it support}  of $f$, denoted by $\cA(f)$.  Thus
we may write
\begin{displaymath}
f=\sum_{a \in \cA(f)} c_{a}t^{a}, \text{ where } t^{a}=t_{1}^{a_{1}} \dotsm t_{m}^{a_{m}},\ c_{a} \in \K.
\end{displaymath} 

The {\it Newton polytope} $P(f)$  is the convex hull of the support of $f$.
It is a convex lattice polytope in $\R^m$. (A polytope is called {\it lattice}  if 
its vertices lie in $\Z^m\subset\R^m$.)
For example, the Newton polytope of $f(t_1,t_2)=t_1^{-1}+2t_1^{-1}t_2-3t_1t_2$ is the triangle
with vertices $(-1,0)$, $(-1,1)$ and $(1,1)$.

Notice that it makes sense to evaluate
Laurent polynomials at points none of whose coordinate is zero, i.e.,
points in the algebraic torus $\T^m=(\K^*)^m$.  Laurent polynomials with a prescribed Newton polytope
are usually called {\it sparse polynomials} to emphasize that, compared
to a generic polynomial of the same degree, it may have only a few monomials (the ones that
correspond to the lattice points in its Newton polytope).

The Newton polytope plays the role of the degree for a sparse polynomial. Note that for
any two sparse polynomials $f,g$ we have $P(fg)=P(f)+P(g)$, just as for usual degrees. 
The sum here is the {\it Minkowski sum} of the polytopes, which is the set of all sums $p_1+p_2$ for all
pairs $p_1\in P(f)$ and $p_2\in P(g)$, and turns out to be again a polytope.
Therefore, factorizations of a sparse polynomial are related to Minkowski sum decompositions
of its Newton polytope. We will see in \rs{TC} how this relation helps to estimate the number
of solutions to $f=0$ over a finite field in terms of the Newton polytope $P(f)$. 

Here is a bit of terminology. We say a lattice segment in $\R^m$ is {\it primitive}
if it contains exactly two lattice points. We say a lattice simplex $\R^m$ is {\it unimodular } 
if it contains exactly $m+1$ lattice points. We say a lattice triangle in $\R^2$ is {\it exceptional}
if it contains exactly three boundary lattice points and one interior lattice point.

\section{Toric Codes}\label{S:TC} 
Let  $\{p_1,\dots,p_n\}$ be the set of all points
in the algebraic  torus $\T^m=(\F_q^*)^m$ in some linear order.
Fix a lattice polytope $P\subset\R^m$ and 
let $\cL(P)$ be the finite-dimensional space of Laurent polynomials over $\F_q$
whose support is contained in $P$: 
\begin{equation}\label{e:L}
\cL(P)=\spn_{\F_q}\{t^a\ |\ a\in P\cap\Z^m\}.
\end{equation}
We have the following  {\it evaluation map}
\begin{equation}\label{e:eval}
ev_{\T^m}:\cL(P)\to\F_q^{|Z|},\quad f\mapsto (f(p_1),\dots,f(p_n)).
\end{equation}
The image of $ev_{\T^m}$ is called the {\it toric code} and is denoted by $\cC_{P}$.

\begin{Rem}  One may regard toric codes as a multivariate generalization of the Reed--Solomon codes. 
Indeed, if $m=1$ and $P$ is the lattice segment $[0,\ell]$ the toric code $\cC_P$
coincides with the Reed--Solomon code with parameters $[q-1,\ell+1,q-1-\ell]_q$.
\end{Rem}

Clearly, the block length $n$ of $\cC_{P}$ equals $(q-1)^m$, the size of $\T^m$. In \cite{Ru}
D. Ruano showed that the dimension $k$ of $\cC_{P}$ equals the number of
lattice points of $P$ if no two of them are congruent modulo $(\Z_{q-1})^m$.
In particular, this is true if we assume that $P$ is contained in the cube
$K_q^m=[0,q-2]^m$. The main problem we are concerned with is how to compute or estimate the
minimum distance $d=d(\cC_P)$.

We will start with some explicit results. J.~Little and R.~Schwarz in \cite{LSch}  computed the minimum distance 
of $\cC_{P}$ in the case of $P=\ell\D_m$, the standard $m$-simplex of side length $\ell$ and 
$P=\Pi_{\ell_1,\dots,\ell_m}$, the product of $m$ segments $[0,\ell_1]\times\cdots \times[0, \ell_m]$:
$$
d(\cC_{\ell\D_m})=(q-1)^{m-1}(q-1-\ell),\quad d(\cC_{\Pi_{\ell_1,\dots,\ell_m}})=\prod_{i=1}^m(q-1-\ell_i).
$$
It turned out that this is an instance of a general phenomenon. In the following theorem
we describe how the minimum distance behaves under basic operations on lattice
polytopes (see \cite{SoSo2} for details).

\begin{Th}\cite{SoSo2}\label{T:SoSo2}
\begin{enumerate}
\item Let $P\subseteq K_q^{m_1}$ and $Q\subseteq K_q^{m_2}$ be lattice polytopes. Then  
$$d(\cC_{P\times Q})=d(\cC_P)\,d(\cC_{Q}).$$
\item Let $Q$ be a lattice polytope of $\dim Q\geq 1$, and let $\{kQ\ |\ 0\leq k\leq N\}$ 
be a sequence of $k$-dilates of $Q$,  contained in $K_{q}^m$. Let $\cP(Q)$ be the pyramid over $Q$, i.e. 
the convex hull in $\R^{m+1}$ of the set $\{(x,0)\ |\ x\in Q\}\cup\{e_{m+1}\}$. Then 
$$d(\cC_{k\cP(Q)})=(q-1)\,d(\cC_{kQ}).$$
\end{enumerate}
\end{Th}

Using this result one can compute the minimum distance explicitly for 
a large class of polytopes obtained from a lattice segment by taking the
direct product  or constructing a pyramid and dilating. In particular,
Umana and Velasco \cite{UmVe} used this to compute the minimum
distance for toric codes on {\it degree one} polytopes.
In \rs{GTC} we generalize this theorem to generalized toric codes.

Next we turn to the case of arbitrary polytopes. The situation is far from being understood
even in the case of polytopes of small dimension. In dimensions two and three we have lower bounds 
on the minimum distance $d(\cC_P)$ in terms of what is called the  Minkowski length of~$P$.
Here is the definition.

\begin{Def}\label{D:ML} Let $P$ be a lattice polytope in $\R^m$.
The {\it  Minkowski length} of $P$ is the maximum 
number of lattice polytopes of positive dimension whose Minkowski
sum is contained in $P$:
 $$L(P)=\max\{\ell\, | \, Q_1+\dots+Q_\ell\subseteq P, \dim Q_i>0\}.$$
A Minkowski decomposition of $Q$ into $L(P)$ summands
of positive dimension will be referred to as  a {\it maximal decomposition in P}
and $Q$ will be called {\it maximal}.
\end{Def}

It is not hard to see that there are only finitely many lattice polytopes $Q$ contained in $P$
and there are only finitely many possible decompositions of $Q$
into the Minkowski sum of lattice polytopes of positive dimension, so the number
$L(P)$ is well-defined. Moreover, it is easy to see that in the definition
of $L(P)$ one may assume that the $Q_i$ are lattice segments. 

Recall from \rs{Prel} that a factorization of a sparse polynomial corresponds 
to Minkowski sum decomposition of its Newton polytope. Therefore,
the  Minkowski length is the geometric invariant of $P$ which
describes the largest possible number of factors in factorizations of polynomials $f\in\cL(P)$.

Consider the case  $m=2$. One can use the Hasse--Weil bound  
to estimate the number of zeroes  in  $\T^2$ of absolutely irreducible factors of  $f\in\cL(P)$.
Little and Schenck in \cite{LSc} used this bound to show that the more factors
$f$ has, the more it has zeroes in $\T^2$, provided $q$ is large enough.
It turns out that if $f\in \cL(P)$ has a factorization with the largest number of factors
then the Newton polytope of each factor is either a primitive segment, or a unimodular
triangle, or an exceptional triangle, see \cite{SoSo}.
Moreover, we have the following lower bound for the minimum distance of $\cC_P$.

\begin{Th}\cite{SoSo}\label{T:SoSo}
Let $P$ be a lattice polygon of  Minkowski length $L$. There is an
explicit function $\alpha(P)$ such that for all $q\geq\alpha(P)$ we have
$$d({\cC}_P)\geq (q-1)(q-1-L)-{(2\sqrt{q}-1)}.$$
Moreover, the term $2\sqrt{q}-1$ may be omitted if no maximal
decomposition of $P$ contains an exceptional triangle.
\end{Th}

There is a natural action of the
isomorphism group $\AGL(m,\Z)$ of the lattice $\Z^m$ on the space of lattice polytopes, under
which $L(P)$ is invariant. The group $\AGL(m,\Z)$ consists of translations by a lattice vector
and integer linear non-degenerate transformations, called unimodular transformations.
Let $P$ and $P'$ be $\AGL(m,\Z)$-equivalent. Then the corresponding
toric codes $\cC_P$ and $\cC_{P'}$ are monomially equivalent \cite{LSch}
(although the opposite is not true, see \cite{Luo} for a counterexample). This means
that for the purpose of coding theory it is enough to consider lattice polytopes up 
to $\AGL(m,\Z)$-equivalence. 

Returning to \rd{ML}, note that each summand in a maximal decomposition
has $L(Q_i)=1$. Such polytopes are called {\it strongly indecomposable} and they
 play an important role in estimating the minimum distance 
$d(\cC_P)$, see \cite{SoSo}, as well as  \cite[Chapter 2]{Whit}.

In dimension $m=2$ there are exactly three strongly indecomposable polytopes up
to $\AGL(m,\Z)$-equivalence: the unit segment, the unit triangle, and the exceptional
triangle, see \rf{3}. 

 \begin{figure}[h]
\includegraphics[scale=.55]{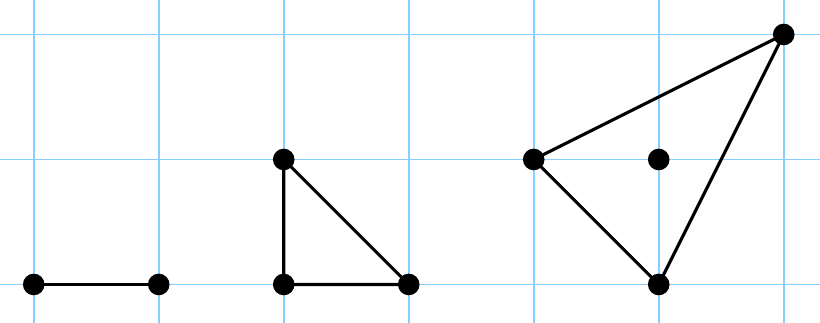}
\caption{Strongly indecomposable polytopes up to $\GL(2,\Z)$-equivalence.}
\label{F:3}
\end{figure}

Note that the latter has the largest number of 
lattice points, which is four. The following theorem is a generalization of this fact,
which was discovered by I. Barnett, B. Fulan, C. Quinn, and J. Soprunova \cite{REU11}. 
For the sake of completeness we include the proof.

\begin{Th}\label{T:2^m} Let $Q\subset \R^m$ be strongly indecomposable. Then the number of lattice points
in $Q$ is at most $2^m$. Moreover, there exist  strongly indecomposable polytopes with
exactly  $2^m$ lattice points.
\end{Th}
 
 \begin{pf} For the first part, consider the lattice points of $Q$ modulo $(\Z/2\Z)^m$. If
 $Q$ has more than $2^m$ lattice points then there exists distinct lattice points $a,b\in Q\cap\Z^m$ which
 coincide modulo $(\Z/2\Z)^m$. Then the lattice segment $[a,b]\subset Q$ must contain at least one interior
 lattice point, hence, decomposes into lattice segments. This contradicts the assumption that $L(Q)=1$.
 
 The construction of $Q$ for which the bound is attained is by induction on $m$. We start with the
 exceptional triangle in $\R^2$.
 After a unimodular transformation we may assume that it contains no horizontal lattice segments,
 i.e. segments whose direction vector has zero first coordinate.
 We will call the direction vector of a lattice segment in a polytope $P$ simply a {\it direction vector in $P$}.
 
 Assume that $P\subset\R^m$ is a strongly indecomposable polytope with $2^m$ lattice points,
 such that no direction vector in $P$ has zero first coordinate. Let $k$ be the largest
 first coordinate of all direction vectors in $P$. There is a unimodular transformation $\alpha\in\GL(m,\Z)$
 such that every direction vector in $\alpha(P)$ has the first coordinate greater than $k$. For example,
 we can take $\alpha=\alpha_2\oplus id_{m-2}$, where $\alpha_2$ has matrix
 $\left[\begin{matrix}a & 1 \\ a-1 & 1\end{matrix}\right]$ with large enough $a$. 
 
 Finally, let $P'$ be the convex hull of $P\times\{0\}\cup\alpha(P)\times\{1\}$ in $\R^{m+1}$.
 To show that $P'$ is strongly indecomposable it is enough to show that  there
 are no lattice segments of length more than one connecting a point in $P$ and a point in $\alpha(P)$,
 and there are no lattice parallelograms with two vertices in $P$ and two vertices in $\alpha(P)$.
 The former is clear since all lattice points in $P'$ are distinct modulo $(\Z/2\Z)^{m+1}$.
 The latter follows from the fact that the first coordinate of every direction vector in $\alpha(P)$ is  
 greater than the first coordinate of any direction vector in $P$.
  \end{pf}

There has been recent progress in understanding the structure of polytopes with $L(P)=1$
in higher dimensions. In particular, new results have been obtained about 3-dimensional lattice polytopes
and longest Minkowski sum decompositions of their subpolytopes \cite{BGSW}. 
As for the bounds in \rt{SoSo}, a similar approach was taken in \cite{Whit} for 3-dimensional toric codes. The author gives an algorithmic way of obtaining lower bound for the minimum distance, but
one still hopes for more explicit bounds than the ones in \cite{Whit}.

Classifying polytopes of  Minkowski length larger than one is not easy even in dimension $m=2$.
In \rf{16} we present 16 classes of lattice polygons of  Minkowski length two. The proof that these
are all of them is not hard, but tedious, so we do not include it here.

 \begin{figure}[h]
\includegraphics[scale=.7]{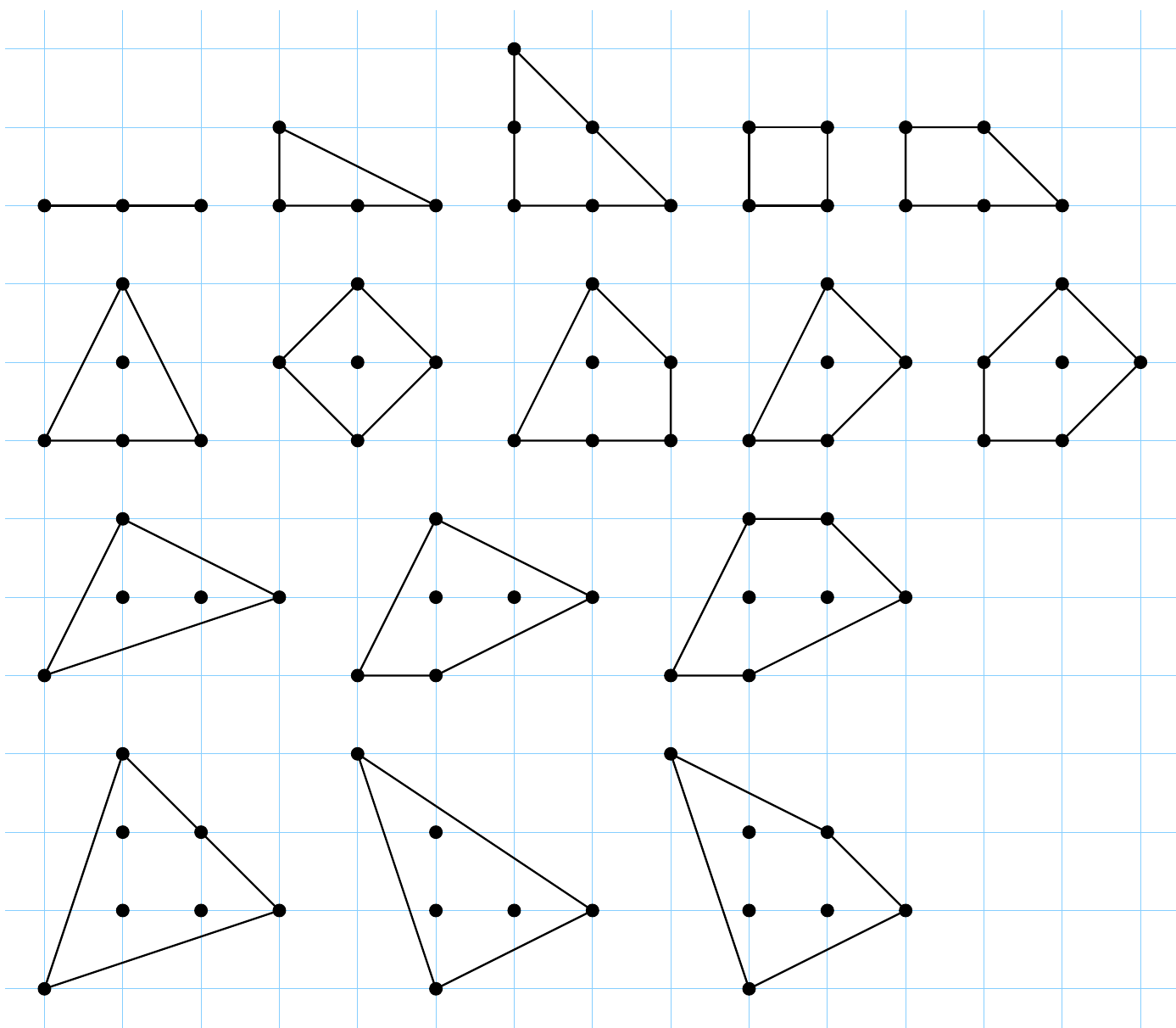}
\caption{The sixteen polytopes with $L(P)=2$  up to $\GL(2,\Z)$-equivalence.}
\label{F:16}
\end{figure}

It does not seem feasible to classify polygons with $L(P)\geq 3$ by hand. Recall that 
the dimension of a toric code equals the number
of lattice points in $P$. Thus,  a more important question is the following: Given $\ell$, what could be the largest number of lattice points in $P$ with $L(P)=\ell$? The naive bound $|P\cap \Z^m|\leq (\ell+1)^m$ which 
follows from considering the lattice points of $P$ modulo  $(\Z/(\ell+1)\Z)^m$, as in the proof
of \rt{2^m}, appears to be too rough.

Suppose $m=2$, so $P$ is a lattice polygon.  From \rf{16} we see that for $\ell=2$ the answer
is $7$. In \cite{Ce}  V.~Cestaro showed that for $\ell=3$ the answer is $9$. For larger $\ell$
the question is open and no better estimate than $(\ell+1)^2$ is currently known.

\section{Generalized Toric Codes}\label{S:GTC}

Generalized toric codes are a natural extension of toric codes. They first 
appeared in the work of D.~Ruano \cite{Ru2} and J.~Little  \cite{Lit11}.
The definition is similar to the one of a toric code, except we allow 
arbitrary configurations of lattice points instead of  the lattice points of a lattice polytope.
More precisely,  let  $S$ be a set of  lattice points in
$\R^m$ contained in the $m$-cube $K_{q}^m$. Similar to \re{L}
we let $\cL(S)$ be the vector space over $\F_q$ of Laurent polynomials 
with support in $S$:
$$\cL(S)=\spn_{\F_q}\{\,t^a\ |\ a\in S\}.$$
The image of the corresponding evaluation map 
\begin{equation}
ev_{\T^m}:\cL(S)\to\F_q^{|Z|},\quad f\mapsto (f(p_1),\dots,f(p_n)).\nonumber
\end{equation}
is called the {\it generalized toric code} $\cC_S$. 
The {weight} of each nonzero codeword equals the number of points
$\xi\in\T^m$ where the corresponding polynomial does not vanish.
We  denote it by $w(f)$. Let $Z(f)$ denote the number of zeroes of $f$ in $\T^m$.
Also  let $Z_S$ denote the maximum number of zeroes 
over all nonzero $f\in\cL(S)$. Obviously, 
\begin{equation}\label{e:def}
Z(f)=(q-1)^m-w(f)\quad\text{ and }\quad Z_S=(q-1)^m-d(S).
\end{equation}

As before, $\cC_S$ is a linear code of block length $n=(q-1)^m$
and dimension $\dim\cC_s=|S|$, the cardinality of $S$. Note that if $P$
is the convex hull of $S$ then 
$$\dim \cC_S\leq \dim \cC_P\quad\text{ and }\quad d(\cC_S)\geq d(\cC_P).$$
The idea is that by omitting just a few lattice points of $P$ one could,
in principle, obtain $S$ for which the minimum distance $d(\cC_S)$
is significantly larger than $d(\cC_P)$. Examples of this phenomenon
were provided by J. Little \cite{Lit11}. At the same time he gave some
evidence that for large $q$ this often does not happen. 

This prompted a search for generalized toric codes with parameters better
than previously known over fields of small size. G. Brown and A. Kasprzyk \cite{BrKasp13, BrKasp-seven}
used an exhaustive search of lattice polygons and lattice point configurations
contained in $K_q^2$ for $q$ up to 8. They were able to find a new
toric code champion and seven new generalized toric code champions.

Below we present two generalized toric code champions found independently of 
G. Brown and A. Kasprzyk. Our approach was to start with small configurations
which produce best known codes and then extend them by adding a lattice point
one by one checking if we obtained a new champion. We used the computer software
Magma \cite{Magma} for our calculations. This way we 
found a $[49, 13, 27]$-code over $\F_8$ by constructing $S$ as in \rf{champs}
on the left. 

 \begin{figure}[h]
\includegraphics[scale=.37]{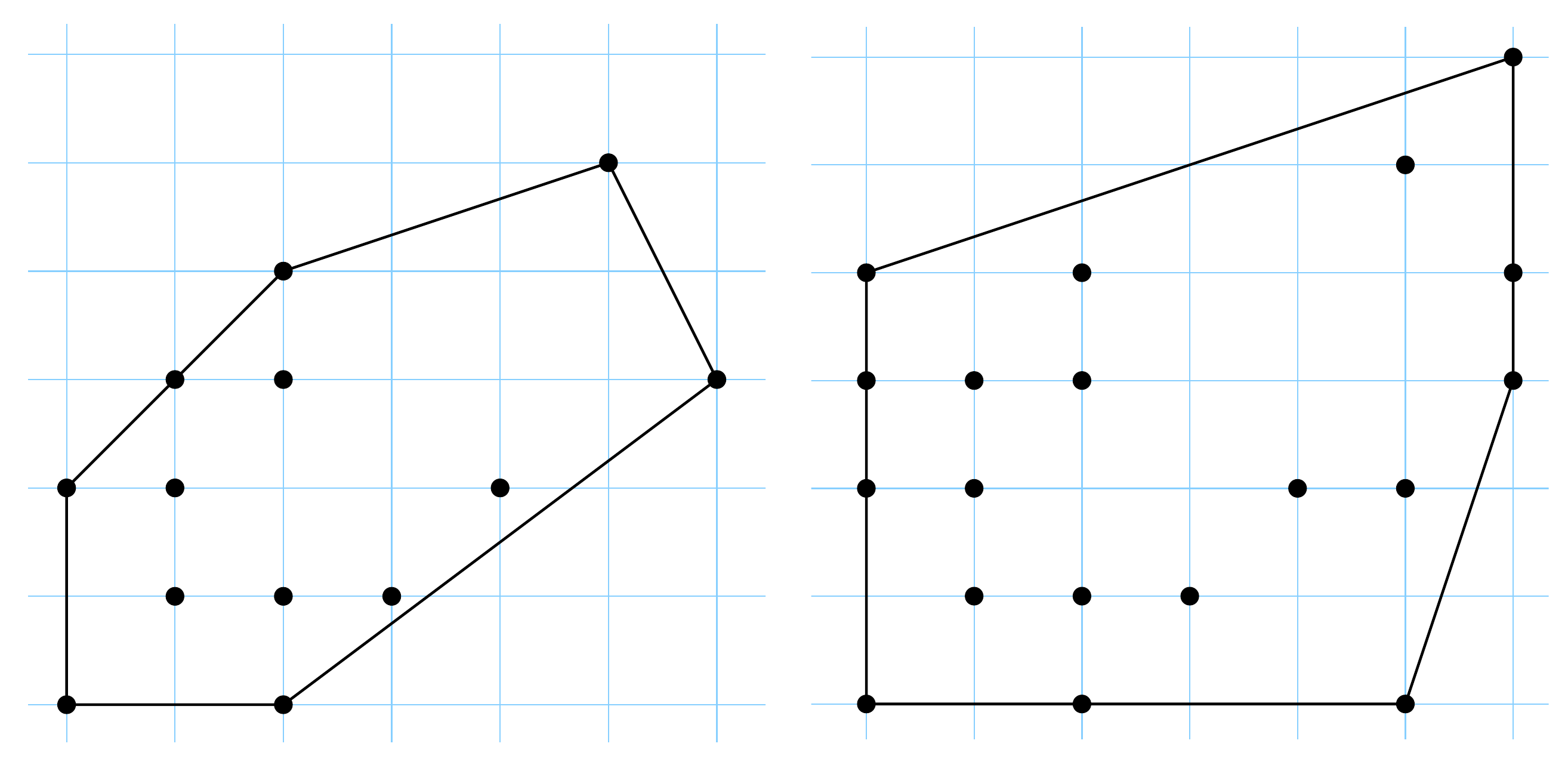}
\caption{Two lattice configurations producing a $[49, 13, 27]$- and $[49, 19, 21]$-code over $\F_8$.}
\label{F:champs}
\end{figure}

Then we were able to extend it to a larger configuration as in the right of 
 \rf{champs}, which produced a $[49, 19, 21]$-code over $\F_8$.
As pointed out by  Markus Grassl in a private communication, by 
omitting the point $(1,2)$ in $S$ one obtains a subcode with parameters $[49,12,28]$.
Applying Construction X to this pair of codes (see \cite{Gra}), one obtains a $[50,13,28]$-code
over $\F_8$, which is another champion.

We finish with a new general lower bound for the minimum distance 
of generalized toric codes.  The bound is inductive in a sense that
it uses the codes from the fibers and the images of a projection of $S$
onto a coordinate subspace. As a corollary we get a generalization
of \rt{SoSo2} to generalized toric codes.

Let $S\subseteq K_{q}^m$ be a set of lattice points.
Choose a coordinate subspace $Y\subseteq \R^m$ and 
let $\pi:\R^m\to Y$ be the corresponding projection.
 For every $a\in\pi(S)$ let $S_a$
denote the fiber $S_a=S\,\cap\,\pi^{-1}(a)$.  

\begin{Th}\label{T:induction}
Let $S$ be a set of lattice points in $K_{q}^m$ and $\pi:\R^m\to Y$ a projection
onto a coordinate subspace. 
 Then 
$$d(S)\geq \min_{S'\subseteq \pi(S)}\left(d(S')\max_{a\in S'}d(S_a)\right).$$
\end{Th}

\begin{pf} We may assume that $\pi:\R^m\to Y$ is the projection
onto the last $m-k$ coordinates. Furthermore, we use
 $(x,y)=(x_1,\dots,x_k,y_1,\dots,y_{m-k})$ 
to denote coordinates in $\T^m=\T^k\times \T^{m-k}$.

 Consider an arbitrary nonzero  $f\in\cL(S)$ with support $\cA(f)$, and 
let  $S'$ denote the projection $S'=\pi(\cA(f))$.
We have $\cA(f)\subseteq \cup_{a\in S'}S_a$,
hence, we can write $f$ as a linear combination of
monomials $y^a$ for  $a\in S'$ with coefficients $f_a$ that 
are nonzero polynomials in $\cL(S_a)$:
\begin{equation}\label{e:goodform2}
f(x,y)=\sum_{a\in S'}f_a(x)y^a.
\end{equation}

Given a point $\xi=(\xi_1,\dots,\xi_k)\in(\F_q^*)^k$ let
$L_\xi$ be the coset of the subtorus $\{1\}\times(\F_q^*)^{m-k}$ containing $\xi$, i.e.
$$L_\xi=\{(\xi,y)\ |\ y\in (\F_q^*)^{m-k}\}.$$

Note that on every $L_\xi$ where $f$ is identically zero,
$f$ has exactly $(q-1)^{m-k}$ zeroes, and on every $L_{\xi}$ where $f$ is not
identically zero, it has at most $Z_{S'}$ zeroes, since the (nonzero) polynomial $f(\xi,y)$
lies in $\cL(S')$.

Then the number of zeroes of $f$ in $\T^m$  is bounded by
\begin{equation}\label{e:long1}
Z(f)\leq (q-1)^{m-k}N+Z_{S'}\left((q-1)^{k}-N\right),
\end{equation}
where $N$ is the number of the cosets $L_{\xi}$ where $f$ is identically zero.
Substituting $Z_{S'}=(q-1)^{m-k}-d(S')$ (see \re{def}) and simplifying we obtain
$$Z(f)\leq (q-1)^m-d(S')\left((q-1)^k-N\right),$$
or, simply,
\begin{equation}\label{e:long2}
w(f)\geq d(S')\left((q-1)^k-N\right).
\end{equation}

Notice that $N$ is, in fact, the number of common zeroes of the $f_a$ in
$(\F_q^*)^{k}$, and is at most the number of zeroes of each $f_a$. 
Therefore,
$$N\leq \min_{a\in S'} Z(f_a)\leq (q-1)^{k}-\max_{a\in S'}d(S_a).$$ 
Now \re{long2} implies
\begin{equation}\nonumber
w(f)\geq d(S')\max_{a\in S'}d(S_a).
\end{equation}
Notice that the right hand side depends only on the projection of the support of $f$,
so it remains to take the minimum over all subsets $S'\subseteq \pi(S)$ and
the statement of the theorem follows.
\end{pf} 

Our first application of the inductive formula is a generalization of \rt{SoSo2}, part (1).

\begin{Cor} Suppose $S=S_1\times S_2\subset\R^{m_1}\times\R^{m_2}$ for some lattice sets
 $S_i\subseteq K_q^{m_i}\cap\Z^{m_i}$, $i=1,2$. Then $d(S)=d({S_1})d({S_2})$.
\end{Cor}

\begin{pf}
Consider the projection $\pi:\R^{m_1}\times\R^{m_2}\to \R^{m_2}$. Then $\pi(S)=S_2$.
As every fiber $S_a$ equals a lattice translate of $S_1$, for $a\in S_2$, by \rt{induction} we have 
$$d(S)\geq \min_{S'\subseteq S_2}\left(d(S')d(S_1)\right)=d(S_1) \min_{S'\subseteq S_2}d(S').$$
It is clear that if $S'\subseteq S_2$ then $d(S')\geq d(S_2)$. Therefore, the above minimum equals $d(S_2)$.

Conversely, let $f_i\in\cL(S_i)$ for $i=1,2$ be polynomials with the minimum weight. We have
$d(S_i)=w(f_i)=(q-1)^{m_i}-Z(f_i)$, where $Z(f_i)$ is the number of zeroes of $f_i$ in $\T^{m_i}$.
Then, by the inclusion-exclusion principle, the polynomial $f=f_1f_2$ has 
$$(q-1)^{m_2}Z(f_1)+(q-1)^{m_1}Z(f_2)-Z(f_1)Z(f_2)$$ zeroes in $\T^{m_1}\times\T^{m_2}$.
This implies that its weight equals
$$w(f)=(q-1)^{m_1+m_2}-(q-1)^{m_2}Z(f_1)-(q-1)^{m_1}Z(f_2)+Z(f_1)Z(f_2)=w(f_1)w(f_2).$$
Therefore, $d(S)\leq d(S_1)d(S_2)$, and we are done.
\end{pf}

\begin{Cor}\label{C:last} 
Let $\pi_m:\R^m\to\R$ be the projection to the last coordinate and suppose 
$\pi_m(S)=\{0,1,\dots, \ell\}$. If  $d(S_{0})\leq d(S_{1})\leq\dots\leq d(S_{\ell})$ then
$$d(S)\geq\min_{0\leq i\leq \ell}(q-1-i)d(S_i).$$
\end{Cor}

\begin{pf} Indeed, consider $S'\subset \pi_m(S)$ and let $i$ be the length of the convex hull of $S'$.
On one hand we have $d(S')\geq (q-1-i)$. On the other hand, since  $d(S_{0})\leq d(S_{1})\leq\dots\leq d(S_{\ell})$,
when finding the minimum over all $S'$ it is enough to consider only those $S'$ that contain~$0$.
In that case $\max_{a\in S'}d(S_a)=d(S_i)$ and the statement follows from \rt{induction}.
\end{pf}

To connect this result to the second part of \rt{SoSo2}, we will need an extra assumption 
on the configuration $S$. First, we have the following proposition. Its proof is similar to the one of \cite[Proposition 2.2]{SoSo2}

\begin{Prop}\label{P:last} Let $S,S'$ be lattice sets in $K_q^m$ and 
$T$ the set of lattice points of a lattice segment. If $S+T\subseteq S'$ (up to a lattice translation)
then  $(q-1)d(S')\leq (q-|T|)d(S)$.
\end{Prop}

\begin{pf} After a unimodular transformation we may assume that $S+T\subseteq S'$, and $T$ is
the set of lattice points of the segment $[0,ke_1]$, where $e_1$ is the first basis vector
and $k=|T|-1$ is the length of the segment. 

Let $g\in\cL(S)$ be a polynomial with $Z(g)=Z_S$. Then for any $\xi_1,\dots, \xi_k\in\F_q^*$ the polynomial
$$f(x)=g(x)\prod_{j=1}^k(x_1-\xi_j)$$
belongs to $\cL(S+T)\subseteq \cL(S')$. 
By the inclusion-exclusion formula we have
$$Z(f)=Z(g)+k(q-1)^{m-1}-\sum_{j=1}^k Z(g|_{x_1=\xi_j}).$$

Since $\T^m$ is the union of $q-1$ subtori given by $x_1=\xi$, for $\xi\in\F_q^*$, we have
$Z(g)=\sum_{\xi\in\F_q^*}Z(g|_{x_1=\xi})$. Choose 
$\xi_1,\dots, \xi_k\in\F_q^*$ so that $\{Z(g|_{x_1=\xi_j})\ |\ j=1,\dots,k\}$ are the $k$ smallest integers
 among the $q-1$ integers $\{Z(g|_{x_1=\xi})\ |\ \xi\in\F_q^*\}$. Then 
 $$\frac{1}{k}\sum_{j=1}^k Z(g|_{x_1=\xi_j})\leq \frac{Z(g)}{q-1}.$$
Therefore, we obtain
$$Z_{S'}\geq Z(f)\geq Z(g)+k(q-1)^{m-1}-\frac{k}{q-1}Z(g)$$
Replacing $Z(g)$ with $Z_S$ and using $Z_S=(q-1)^m-d(S)$ we see that the latter inequality is
equivalent to $(q-1)d(S')\leq (q-k-1)d(S)$, as required.
\end{pf}




The following is a generalization of  \rt{SoSo2}, part (2) to generalized toric codes.

\begin{Th} Let $S$ be a lattice set in $\K_q^m$. Let $\pi_m:\R^m\to\R$ be the projection to the last coordinate, 
$\pi_m(S)=\{0,1,\dots, \ell\}$, and $S_0,\dots, S_\ell$ the corresponding fibers. 
Suppose there is a primitive lattice segment $[a,b]$ such that for every $1\leq i\leq \ell$, the set
$S_i+\{a,b\}$  is contained in $S_{i-1}$, up to a lattice translation. Then
$$d(S)= (q-1)d(S_0).$$
\end{Th}

\begin{pf} First, note that in this special situation,
the conditions of \rc{last} are satisfied. Indeed, by
\rp{last}, $(q-1)d(S_{i-1})\leq (q-2)d(S_i)$, so in particular, $d(S_{i-1})\leq d(S_i)$.

Next, we have $S_i+i\{a,b\}\subseteq S_0$ up to a lattice translation. Here
$i\{a,b\}$ (which is the Minkowski sum of $\{a,b\}$ with itself $i$ times)
is the set of lattice points of a lattice segment of length $i$. Thus, by \rp{last},
$$(q-1)S_0\leq (q-1-i)d(S_i),$$ 
for every $0\leq i\leq \ell$. Applying \rc{last}, we obtain
$$d(S)\geq (q-1)d(S_0).$$

Conversely, let $g\in\cL(S_0)$ be a polynomial with $Z(g)=Z_{S_{0}}$. By
definition, $g$ depends only on the first $m-1$ variables. Therefore, it
has $(q-1)Z_{S_0}$ zeroes in $\T^m$. This implies that $Z_S\geq (q-1)Z_{S_0}$, i.e.
$d(S)\leq (q-1)d(S_0)$.

\end{pf}

\section*{Acknowledgments} The first part of this paper is based on a talk given
at Karatekin Mathematics Days 2014 International Mathematics Symposium.
I am grateful to the organizers for inviting me and  to Mesut \c{S}ah\.{i}n, Pinar Celebi Demirarslan,
and G\"{o}khan Demirarslan  for their hospitality.

\end{document}